\documentclass[11pt]{article}

\usepackage{eulervm}
\usepackage{setspace}
\usepackage[pdftex]{graphics}
\usepackage{graphicx}
\usepackage{color}
\usepackage{latexsym}
\usepackage{amsfonts}
\usepackage{amssymb, amsmath}
\usepackage{amsthm}
\usepackage{subcaption}
\usepackage{float}
\usepackage{bm}
\DeclareGraphicsExtensions{.pdf,.png,.jpg}

\usepackage{natbib} 
\usepackage{enumerate}

\usepackage[colorlinks=true, pdfstartview=FitV, linkcolor=blue, citecolor=blue, urlcolor=blue]{hyperref}

\usepackage{tikz}

\usepackage{epsfig}

\usepackage{stata}

\begin{document}
	
	\title{Quantile Regression under Limited Dependent Variable}
	
	\author{Javier Alejo\footnote{IECON-Universidad de la Rep\'ublica, Uruguay. E-mail: javier.alejo@ccee.edu.uy} 
		\and 
		Gabriel Montes-Rojas\footnote{Universidad de Buenos Aires and CONICET, Argentina. e-mail: gabriel.montes@fce.uba.ar}
	}
	
	\maketitle
	
	\begin{abstract}
		A new Stata command, \texttt{ldvqreg}, is developed to estimate quantile regression models for the cases of censored (with lower and/or upper censoring) and binary dependent variables. The estimators are implemented using a smoothed version of the quantile regression objective function. Simulation exercises show that it correctly estimates the parameters and it should be implemented instead of the available quantile regression methods when censoring is present. An empirical application to women's labor supply in Uruguay is considered.
		
		\vspace{3mm}
		
		\vspace{3mm}
	\end{abstract}
	
	\onehalfspacing 
	\newpage

\section{Introduction}

Quantile regression (QR) is an important method for modeling heterogeneous effects. It allows to study generalized regression models by focusing on the quantiles of the conditional distribution of an outcome variable, controlling for observable covariates.

Censoring is an important feature in applied models. Mean-based regression models are analyzed in many ways with respect to censoring and truncation of the dependent variable. Typical examples are the Tobit-type models (see e.g. \texttt{tobit}) and binary regression models (see e.g. \texttt{logit} and \texttt{probit}). 

This paper provides an integrated command that allows to estimate censored quantile regression models. First, it considers lower and upper censoring by developing the counterpart of a mean-based Tobit model for QR. Censored QR has been studied in several papers, including to cite a few \cite{Powell84,Powell86}, \cite{Buchinsky91}, \cite{ChernozhukovHong02}.
Second, it provides a model for semi-parametric binary regression. This has been studied in the seminal papers of \cite{Manski85,Manski91} and \cite{Kordas06} among others. These two options allow to apply QR models in a wide variety of frameworks where the dependent variable is affected.

The key characteristic of these estimators is that quantile models are invariant to monotone non-decreasing transformations. Thus the conditional effects of covariates can be recovered from the transformed model. Moreover, one important common feature is that the usual algorithm for QR involves linear programming (see for instance the \texttt{qreg} command and the related packages). Recent applications of QR emphasize that the estimators are improved in both asymptotics and numerical accuracy if the objective function is replaced by a smooth counterpart, see e.g. \cite{Horowitz92}, \cite{Kordas06}, \cite{Kaplan17} and \cite{dCGKL19}. The proposed estimator follows this strategy.

Alternative procedures have been developed to accommodate for censoring in QR. \cite{sg153} (\texttt{clad} command) uses \cite{Buchinsky91} and \cite{BuchinskyHahn98} algorithm to compute QR with censoring. \cite{Baker13} uses Monte Carlo integration techniques to accomplish the same. Our command contributes to this list.

The remainder of the paper is organized as follows. Section \ref{models} presents the censored and binary QR models. Section \ref{ldvqreg} summarizes the \texttt{STATA} command features. Section \ref{examples} present numerical simulations and an empirical application to women's labor supply in Uruguay. Section \ref{conclusion} concludes.

\section{Limited dependent variable models in quantile regression}\label{models}

Consider the following conditional $\tau$-quantile model,
$$Q_{\tau}(y_{i}^{*}|\boldsymbol x_{i}) = \boldsymbol x_{i}'\boldsymbol\beta(\tau),$$
where $y_{i}^{*}$ is a latent unobservable variable and $x_{i}$ corresponds to observable covariates. Assume that we have another observable variable $y_{i} = h(y_{i}^{*})$, where $h(.)$ is a non-decreasing monotone transformation, which is defined as a censored variable. The main feature in these models is that we observe $y_i$ but not $y_i^*$. However, by a common characteristic in quantile models, the so-called the quantile invariance property implies that $Q_{\tau}[h(y^{*})|\boldsymbol x]=h[Q_{\tau}(y^{*}|\boldsymbol x)]$, then $Q_{\tau}[y|\boldsymbol x]=h[Q_{\tau}(y^{*}|\boldsymbol x)]=h[\boldsymbol x'\boldsymbol\beta(\tau)]$. That is, the observable variable quantile is a transformed version of the original linear model.

This model includes as particular cases the \textit{censored quantile regression} and the \textit{binary dependent variable regression} models. We study these two cases in the following sections.

\subsection{Censored Quantile Regression}

Consider the case where there is an upper censoring  (at value $c_H$) and lower censoring (at value $c_L$, for $c_H>c_L$) of the dependent variable such that

\begin{equation}
y_{i}=\left\{\begin{array}{cc}
c_L 			& 	 si\; y_{i}^{*}<c_L 	\\
y_{i}^{*} 	&	 si\; c_H\geq y_{i}^{*}\geq c_L 	\\
c_H 			& 	 si\; y_{i}^{*}>c_H 	\\
\end{array}
\right.
\end{equation}
For this case $y=h(y^{*}) = min[max(y^{*},c_L),c_H]$, which is a non-decreasing monotone transformation. This model supports the case of no or partial censoring if we consider $c_L=-\infty$ and /or $c_H=+\infty$.

\cite{Powell84,Powell86} proposes to estimate $\boldsymbol \beta(\tau)$ by

\begin{equation}
\boldsymbol{\hat{\beta}}(\tau)=arg\underset{\boldsymbol b\in\mathbb{R}^{K}}{min}\,\,\,n^{-1} \underset{i=1}{\overset{n}{\sum}}\rho_{\tau}\{y_{i} - min[max(\boldsymbol{x}_{i}'\boldsymbol{b},c_L),c_H]\}
\end{equation}
where $\rho_{\tau}(u)$ is the check function as in \cite{KoenkerBassett78}. This is defined as the censored quantile regression (CQR) model.

\subsection{Binary Quantile Regression}

Consider now the case of a binary dependent variable model with

\begin{equation}
y_{i}=\left\{\begin{array}{cc}
0 	& 	si\;y_{i}^{*}\leq0\\
1 	& 	si\;y_{i}^{*}>0
\end{array}
\right.
\end{equation}

Note that this is also a non-decreasing monotone transformation of the dependent variable, $y=h(y^{*}) = 1(y^{*}>0)$. With this idea \cite{Manski75,Manski85} proposes a maximum score estimator based on:

\begin{equation}
\boldsymbol{\hat{\beta}}(\tau)=arg\underset{\boldsymbol{b}:\Vert \boldsymbol{b}\Vert=1}{min}\,\,\,n^{-1}\underset{i=1}{\overset{n}{\sum}}\rho_{\tau}[y_{i}-I\{\boldsymbol{x}_{i}'\boldsymbol{b}\geq0\}].
\end{equation}
which can be written as

\begin{equation}
\boldsymbol{\hat{\beta}}(\tau)=arg\underset{\boldsymbol{b}:\Vert \boldsymbol{b}\Vert=1}{max}\,\,\,n^{-1}\underset{i=1}{\overset{n}{\sum}}[y_{i}-(1-\tau)]I\{\boldsymbol{x}_{i}'\boldsymbol{b}\geq0\}.
\end{equation}
This is defined as the binary quantile regression (BQR) model.

\subsection{Smoothed Quantile Regression}

Both, the censored and the binary cases, share a common feature. The theoretical formulation provides estimators with a proper characterization, identification and which are consistent and asymptotically normal (see the corresponding cited papers above). However, its numerical performance in applied cases is very poor. In particular, the maximization problem does not provide satisfactory numerical solutions in general. This is a common feature in some variants of QR models, and the consensus in the literature is to provide smooth objective functions alternatives. 

For our purposes, \cite{Horowitz92} and \cite{Kordas06} propose to smooth the objective function by using 

\begin{equation*}
K(\boldsymbol{x}_{i}'\boldsymbol{b}/h_{n}),
\end{equation*}
the integral of a kernel function with $h_n$ bandwidth instead of  $I\{\boldsymbol{x}_{i}'\boldsymbol{b}\geq0\}$. This provides remarkable improvements in applied cases. Our proposed estimator follows this strategy.  The smoothing function we use is the cumulative distribution of a Gaussian kernel. 

Both in the case of censored and binary regression models we use the same heuristic rule for the choice of bandwidth: the same formula used by \texttt{STATA} to estimate densities with kernel functions. This is,

\begin{equation*}
h_n = \frac{0.9\cdot\hat{\sigma}_u}{n^{1/5}}
\end{equation*}
where $\hat{\sigma}_u$ is an estimate of the standard deviation of the latent variable conditional distribution. 
In the case of censored data we use the $\hat{\sigma}_u$ estimated by the Tobit model, while in the case of the binary data we set $\hat{\sigma}_u=1$ (the usual normalization of this parameter in the Probit model).

\subsection{Prediction of Censored Quantiles and Probabilities}

An important issue in censored models is the appropriate prediction exercise.

Following \cite{Kordas06}, for the BQR model, we consider the probability of $y=1$, which corresponds to $y^{*}>0$. This can be estimated by computing $\boldsymbol{x}'\boldsymbol{\beta}(U)>0$ where $U \sim U(0,1)$. Given that for the binary case $P(y=1|\boldsymbol{x})=E(y|\boldsymbol{x})$, then $P(y=1|\boldsymbol{x})=E[1(\boldsymbol{x}'\boldsymbol{\beta}(U)>0)|x]$. Therefore,

\begin{equation}
P(y=1|\boldsymbol{x})=\int_{0}^{1}I\{\boldsymbol{x}'\boldsymbol{\beta}(\tau)>0\}d\tau.    
\end{equation}
Then, this can be estimated by a grid of quantile indexes $\{\tau_{1},\tau_{2},...,\tau_{m}\}$ by computing

\begin{equation}\label{eq:phat}
\hat{P}(y=1|\boldsymbol{x}_{i})=m^{-1}\underset{j=1}{\overset{m}{\sum}}I\{\boldsymbol{x}_{i}'\boldsymbol{\hat{\beta}}(\tau_{j})>0\},
\end{equation}
where $\boldsymbol{\hat{\beta}}(\tau)$ is the corresponding BQR estimator. 
An smoothed version of \eqref{eq:phat} replaces the indicator function by the integral of the kernel $K(.)$: 

\begin{equation}\label{eq:phat_s}
\hat{P}(y=1|\boldsymbol{x}_{i})=m^{-1}\underset{j=1}{\overset{m}{\sum}}K[\boldsymbol{x}_{i}'\boldsymbol{\hat{\beta}}(\tau_{j})/h_{n}].
\end{equation}









Equations \eqref{eq:phat} and \eqref{eq:phat_s} are used to compute the probability of censoring for the CQR model with  $c_L \leq y \leq c_H$. Similar to the previous example we get

\begin{equation}
\hat{P}(y=c_{L}|\boldsymbol{x}_{i})=\hat{P}(y^{*}<c_{L}|\boldsymbol{x}_{i})=m^{-1}\underset{j=1}{\overset{m}{\sum}}I\{\boldsymbol{x}_{i}'\boldsymbol{\hat{\beta}}(\tau_{j})<c_{L}\}
\end{equation}
and 
\begin{equation}
\hat{P}(y=c_{H}|\boldsymbol{x_{i}})=\hat{P}(y^{*}>c_{H}|\boldsymbol{x_{i}})=m^{-1}\underset{j=1}{\overset{m}{\sum}}I\{\boldsymbol{x}_{i}'\boldsymbol{\hat{\beta}}(\tau_{j})>c_{H}\},
\end{equation}
where $\boldsymbol{\hat{\beta}}(\tau)$ is the CQR coefficient estimate. The smoothed versions replace $I(.)$  by $K(.)$.






Finally, for the CQR model we can compute the prediction for a given censured quantile $\tau$ as
\begin{equation}
\hat{Q}_{\tau}(y|\boldsymbol{x}_{i})=min\{max[\boldsymbol{x}_{i}'\boldsymbol{\hat{\beta}}(\tau);c_{L}];c_{H}\},
\end{equation}
where $\boldsymbol{\hat{\beta}}(\tau)$ are the CQR estimates.



\section{The \texttt{ldvqreg} syntax}\label{ldvqreg}

In this section we present the syntax of the \texttt{ldvqreg} command.    

\subsection{Sintaxis}

The command syntax is:

\begin{stsyntax}
	ldvqreg
	\depvar\
	\optindepvars\ 
	\optif\
	\optin\
	\optional{,
		\underline{q}uantile(\num[\num[\num ...]]) 
		ll(real)
		ul(real)
		reps(string)
		qcen(string)
		pcen(string)
		p1(string)
		\underline{bw}idth(real)
		\underline{pbw}idth(real)
	}
\end{stsyntax}

\subsection{Options}

\texttt{ldvqreg} supports the following options:
\vspace{.3cm}

General:

\texttt{quantile(\num)} 	estimates \texttt{\num} quantile; default is quantile(50)

\texttt{reps(\num)} 	performs \texttt{\num} bootstrap replications; default is reps(50)
\vspace{.3cm}

Censoring:

\texttt{ll(\num)} 	left-censoring limit

\texttt{ul(\num)} 	right-censoring limit

\texttt{qcen(newvar)} 	stores predicted censored quantiles in \texttt{newvar\underline{ }q\num}

\texttt{pcen(newvar)} 	stores censorship probability in \texttt{newvar} and \texttt{newvar\underline{ }s} (smoothed)
\vspace{.05cm}

Binary data:

\texttt{p1(newvar)} 	stores probability of $depvar=1$ in \texttt{newvar} and \texttt{newvar\underline{ }s} (smoothed)
\vspace{.05cm}

Smoothing:

\texttt{bwidth(real)} 	specifies the bandwidth to smooth the target function

\texttt{pbwidth(real)} 	specifies the bandwidth to smooth the predicted probabilities

\vspace{.5cm}

If \texttt{ll(\num)} and \texttt{ul(\num)} are not specified and the dependent variable is a dummy variable (which is automatically checked), then the command runs a binary QR. If \texttt{ll(\num)} and \texttt{ul(\num)} are not specified but the dependent variable is not a dummy, then the command runs a smoothed QR model.



\subsection{Saved results}

\texttt{ldvqreg} stores the following results in \stcmd{e()}:

\begin{stresults}
	
	\stresultsgroup{Scalars} \\
	\stcmd{e(N)} & number of observations.\\
	\stcmd{e(reps)} & number of replications.\\
	\stcmd{e(bwidth)} & bandwidth.\\
	
	\stresultsgroup{Macros} \\
	\stcmd{e(title)} & Censored or Binary model.\\
	\stcmd{e(vcetype)} & title used to label Std. Err.\\
	\stcmd{e(properties)} & b V.\\
	\stcmd{e(depvar)} & name of dependent variable.\\
	
	\stresultsgroup{Matrices} \\
	\stcmd{e(b)} & coefficients' vector.\\
	\stcmd{e(V)} & bootstrap variance matrix.\\
	
	\stresultsgroup{Functions} \\
	\stcmd{e(sample)} & marks estimation sample.\\
	
\end{stresults}

\section{Examples}\label{examples}

\subsection{Example 1: Simulations}

\subsubsection{Censoring}

We start with a simple simulation to show that the Tobit model can be biased if the homogeneity of the conditional distribution is not satisfied. This can be achieved by  a location-scale model, which is a typical model in QR to generate heterogeneity in the regression coefficients. In turn, this motivates the necessity of using CQR models rather than the mean-based Tobit.


\begin{stlog}
	. set seed 321
	
	. set obs 1000
	number of observations (_N) was 0, now 1,000
	
	. 
	. gen x = runiform()
	
	. gen y = -1/3 + x + x*rnormal()/3
	
	. gen y_c = max(y,0)
	
\end{stlog}

\begin{figure}
	\caption{A comparison of Tobit and censored quantile regression models}
	\label{fig:figure1}
	\centering
	\includegraphics[scale=0.33]{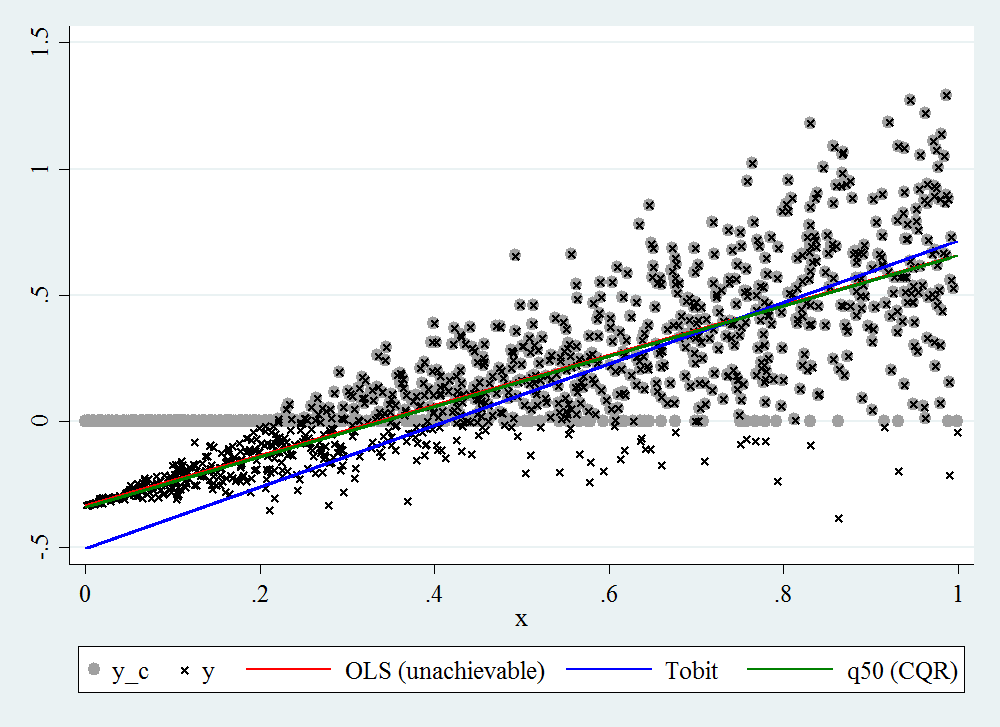}
\end{figure}

In this example the error term is standard Gaussian, but its interaction with \texttt{x} determines that it has conditional heteroscedasticity. Figure \ref{fig:figure1} shows a scatter plot with the latent (unobserved variable) \texttt{y} (using points with an x) and the censored variable \texttt{y\underline{ }c} (with grey circles). The graph also has the OLS estimation of the relation between \texttt{y} and \texttt{x} (which is not feasible as we cannot observe the latent variable), the Tobit estimation that controls for lower censoring at $0$, which assumes homoscedasticity, and the median regression estimate using the CQR estimate with \texttt{ldvqreg} of \texttt{y\underline{ }c} on \texttt{x}, which is distribution free (i.e. it does not require to assume homoscedastic Gaussian errors as in the Tobit model).


Note that the Tobit estimator is clearly different from the true OLS estimate. Nevertheless, the line that corresponds to the CQR estimation is indeed close to the true line. Then, the CQR appears as a useful alternative to the Tobit model when the distributional requirements are not satisfied.


Consider now the following data generating process (DGP). We consider two cases, one with a homoscedastic error structure (\texttt{heter==0}) and another with heteroscedastic error structure (\texttt{heter==1}). The former model has that all the QR coefficients are the same across quantiles, i.e. $\beta(\tau)=\beta,\ \forall \tau$, while the second allows for heterogeneity across quantiles. The generated variable \texttt{y} has no censoring, and \texttt{y\underline{ }c} has lower (0) and upper (1) censoring.

\begin{stlog}
	
	. drop _all
	
	. set seed 321
	
	. set obs 4000
	number of observations (_N) was 0, now 4,000
	
	. 
	. gen heter = (_n>_N/2)
	
	. 
	. gen x = runiform()
	
	. gen y = .
	(4,000 missing values generated)
	
	. replace y = x + rnormal()/3             if heter==0
	(2,000 real changes made)
	
	. replace y = x + (1+x)*rnormal()/3   if heter==1
	(2,000 real changes made)
	
	. gen y_c = min(max(y,0),1)
	
\end{stlog}

We will first compare the available command \texttt{sqreg} with the proposed command \texttt{ldvqreg} in the case where there is no censoring, i.e. without specifying \texttt{ll(\#)} and \texttt{ul(\#)} (thus the command assumes no censoring). This is done to evaluate if the smoothed implementation works. The results below show that the results are very similar, and thus \texttt{ldvqreg} works for the general case.

\begin{stlog}
	
	. sqreg y x if heter==0 , q(20 50 80) reps(100)
	(fitting base model)
	
	Bootstrap replications (100)
	----+--- 1 ---+--- 2 ---+--- 3 ---+--- 4 ---+--- 5 
	..................................................    50
	..................................................   100
	
	Simultaneous quantile regression                    Number of obs =      2,000
	bootstrap(100) SEs                                .20 Pseudo R2 =     0.2514
	.50 Pseudo R2 =     0.2570
	.80 Pseudo R2 =     0.2524
	
	------------------------------------------------------------------------------
	|              Bootstrap
	y |      Coef.   Std. Err.      t    P>|t|     [95
	-------------+----------------------------------------------------------------
	q20          |
	x |   1.017471   .0315822    32.22   0.000     .9555336    1.079409
	_cons |  -.2881128   .0168678   -17.08   0.000    -.3211931   -.2550325
	-------------+----------------------------------------------------------------
	q50          |
	x |   1.030221   .0391766    26.30   0.000     .9533901    1.107053
	_cons |  -.0148498   .0181692    -0.82   0.414    -.0504823    .0207828
	-------------+----------------------------------------------------------------
	q80          |
	x |   1.092272   .0385613    28.33   0.000     1.016648    1.167897
	_cons |   .2442029   .0224929    10.86   0.000     .2000909    .2883149
	------------------------------------------------------------------------------
	
	. test [q20=q50=q80]: x
	
	( 1)  [q20]x - [q50]x = 0
	( 2)  [q20]x - [q80]x = 0
	
	F(  2,  1998) =    1.70
	Prob > F =    0.1837
	
	. ldvqreg y x if heter==0 , q(20 50 80) reps(100)
	(running cqr_est on estimation sample)
	
	Bootstrap replications (100)
	----+--- 1 ---+--- 2 ---+--- 3 ---+--- 4 ---+--- 5 
	..................................................    50
	..................................................   100
	
	Censored quantile regression                    Number of obs     =      2,000
	Replications      =        100
	
	------------------------------------------------------------------------------
	|   Observed   Bootstrap                         Normal-based
	y |      Coef.   Std. Err.      z    P>|z|     [95
	-------------+----------------------------------------------------------------
	q20          |
	x |   1.009058   .0381342    26.46   0.000     .9343168      1.0838
	_cons |  -.2800764   .0225376   -12.43   0.000    -.3242492   -.2359036
	-------------+----------------------------------------------------------------
	q50          |
	x |   1.041249   .0366032    28.45   0.000     .9695078     1.11299
	_cons |  -.0182596   .0212655    -0.86   0.391    -.0599391    .0234199
	-------------+----------------------------------------------------------------
	q80          |
	x |     1.0884   .0344859    31.56   0.000     1.020809    1.155991
	_cons |   .2457266    .019879    12.36   0.000     .2067646    .2846887
	------------------------------------------------------------------------------
	
	. test [q20=q50=q80]: x
	
	( 1)  [q20]x - [q50]x = 0
	( 2)  [q20]x - [q80]x = 0
	
	chi2(  2) =    3.23
	Prob > chi2 =    0.1992
	
	. 
	. sqreg y x if heter==1 , q(20 50 80) reps(100)
	(fitting base model)
	
	Bootstrap replications (100)
	----+--- 1 ---+--- 2 ---+--- 3 ---+--- 4 ---+--- 5 
	..................................................    50
	..................................................   100
	
	Simultaneous quantile regression                    Number of obs =      2,000
	bootstrap(100) SEs                                .20 Pseudo R2 =     0.0922
	.50 Pseudo R2 =     0.1363
	.80 Pseudo R2 =     0.2003
	
	------------------------------------------------------------------------------
	|              Bootstrap
	y |      Coef.   Std. Err.      t    P>|t|     [95
	-------------+----------------------------------------------------------------
	q20          |
	x |   .7749115   .0672129    11.53   0.000     .6430968    .9067262
	_cons |  -.3017343   .0280326   -10.76   0.000    -.3567105   -.2467581
	-------------+----------------------------------------------------------------
	q50          |
	x |    1.01513   .0485833    20.89   0.000     .9198508    1.110409
	_cons |   .0034164   .0240824     0.14   0.887    -.0438129    .0506458
	-------------+----------------------------------------------------------------
	q80          |
	x |   1.309644   .0642509    20.38   0.000     1.183639     1.43565
	_cons |   .2884671   .0252653    11.42   0.000     .2389181    .3380162
	------------------------------------------------------------------------------
	
	. test [q20=q50=q80]: x
	
	( 1)  [q20]x - [q50]x = 0
	( 2)  [q20]x - [q80]x = 0
	
	F(  2,  1998) =   20.09
	Prob > F =    0.0000
	
	. ldvqreg y x if heter==1 , q(20 50 80) reps(100)
	(running cqr_est on estimation sample)
	
	Bootstrap replications (100)
	----+--- 1 ---+--- 2 ---+--- 3 ---+--- 4 ---+--- 5 
	..................................................    50
	..................................................   100
	
	Censored quantile regression                    Number of obs     =      2,000
	Replications      =        100
	
	------------------------------------------------------------------------------
	|   Observed   Bootstrap                         Normal-based
	y |      Coef.   Std. Err.      z    P>|z|     [95
	-------------+----------------------------------------------------------------
	q20          |
	x |   .7489802   .0545416    13.73   0.000     .6420807    .8558797
	_cons |  -.2819161   .0256249   -11.00   0.000    -.3321398   -.2316923
	-------------+----------------------------------------------------------------
	q50          |
	x |   1.019204   .0599742    16.99   0.000     .9016568    1.136751
	_cons |   .0039883   .0260766     0.15   0.878    -.0471208    .0550975
	-------------+----------------------------------------------------------------
	q80          |
	x |    1.32032   .0535042    24.68   0.000     1.215453    1.425186
	_cons |   .2646587   .0221209    11.96   0.000     .2213026    .3080149
	------------------------------------------------------------------------------
	
	. test [q20=q50=q80]: x
	
	( 1)  [q20]x - [q50]x = 0
	( 2)  [q20]x - [q80]x = 0
	
	chi2(  2) =   84.47
	Prob > chi2 =    0.0000

\end{stlog}

Second, we show how to estimate the quantiles with the censored variable using the \texttt{ldvqreg} command. For this we must use the options \texttt{ll(0)} and \texttt{ul(1)} to indicate the lower (0) and upper (1) censoring points that correspond to this case. Note that both results are similar to the \texttt{sqreg} command with \texttt{y} as dependent variable (i.e. no censoring). Therefore, the \texttt{ldvqreg} command allows us to retrieve some of the information about the latent variable distribution.

\begin{stlog}
	
	. ldvqreg y_c x if heter==0 , q(20 50 80) reps(100) ll(0) ul(1)
	(running cqr_est on estimation sample)
	
	Bootstrap replications (100)
	----+--- 1 ---+--- 2 ---+--- 3 ---+--- 4 ---+--- 5 
	..................................................    50
	..................................................   100
	
	Censored quantile regression                    Number of obs     =      2,000
	Replications      =        100
	
	------------------------------------------------------------------------------
	|   Observed   Bootstrap                         Normal-based
	y_c |      Coef.   Std. Err.      z    P>|z|     [95
	-------------+----------------------------------------------------------------
	q20          |
	x |   .9540131   .0689456    13.84   0.000     .8188821    1.089144
	_cons |  -.2385125   .0496892    -4.80   0.000    -.3359016   -.1411234
	-------------+----------------------------------------------------------------
	q50          |
	x |   1.002916   .0240021    41.78   0.000     .9558728    1.049959
	_cons |   .0010923   .0136379     0.08   0.936    -.0256374     .027822
	-------------+----------------------------------------------------------------
	q80          |
	x |   1.090821   .0663381    16.44   0.000     .9608005    1.220841
	_cons |   .2454639   .0247217     9.93   0.000     .1970102    .2939177
	------------------------------------------------------------------------------
	
	. test [q20=q50=q80]: x
	
	( 1)  [q20]x - [q50]x = 0
	( 2)  [q20]x - [q80]x = 0
	
	chi2(  2) =    2.10
	Prob > chi2 =    0.3501
	
	. 
	. ldvqreg y_c x if heter==1 , q(20 50 80) reps(100) ll(0) ul(1)
	(running cqr_est on estimation sample)
	
	Bootstrap replications (100)
	----+--- 1 ---+--- 2 ---+--- 3 ---+--- 4 ---+--- 5 
	..................................................    50
	..................................................   100
	
	Censored quantile regression                    Number of obs     =      2,000
	Replications      =        100
	
	------------------------------------------------------------------------------
	|   Observed   Bootstrap                         Normal-based
	y_c |      Coef.   Std. Err.      z    P>|z|     [95
	-------------+----------------------------------------------------------------
	q20          |
	x |   .5630813   .0856675     6.57   0.000      .395176    .7309865
	_cons |  -.1516944   .0410796    -3.69   0.000     -.232209   -.0711798
	-------------+----------------------------------------------------------------
	q50          |
	x |   .9788213   .0363784    26.91   0.000     .9075208    1.050122
	_cons |     .01849   .0123575     1.50   0.135    -.0057302    .0427102
	-------------+----------------------------------------------------------------
	q80          |
	x |    1.37022   .1536848     8.92   0.000     1.069004    1.671437
	_cons |   .2560525   .0359951     7.11   0.000     .1855034    .3266015
	------------------------------------------------------------------------------
	
	. test [q20=q50=q80]: x
	
	( 1)  [q20]x - [q50]x = 0
	( 2)  [q20]x - [q80]x = 0
	
	chi2(  2) =   32.67
	Prob > chi2 =    0.0000
	
\end{stlog}

Finally, we show that ignoring censoring in the estimation of conditional quantiles introduces a bias by comparing the results of the \texttt{sqreg} and \texttt{ldvqreg} commands using the censored dependent variable.\footnote{This is a generalization of the bias that occurs for censoring in the mean-based model, which can be studied by the Tobit estimator and its comparison to standard OLS.} Since we only compare point estimates, we run only a few replicates of the bootstrap.

\begin{stlog}
	
	. sqreg y_c x, reps(5) q(20 50 80)
	(fitting base model)
	
	Bootstrap replications (5)
	----+--- 1 ---+--- 2 ---+--- 3 ---+--- 4 ---+--- 5 
	.....
	
	Simultaneous quantile regression                    Number of obs =      4,000
	bootstrap(5) SEs                                  .20 Pseudo R2 =     0.1504
	.50 Pseudo R2 =     0.2474
	.80 Pseudo R2 =     0.1968
	
	------------------------------------------------------------------------------
	|              Bootstrap
	y_c |      Coef.   Std. Err.      t    P>|t|     [95
	-------------+----------------------------------------------------------------
	q20          |
	x |   .5620175   .0234281    23.99   0.000     .5160853    .6079496
	_cons |  -.0655659   .0015824   -41.44   0.000    -.0686682   -.0624636
	-------------+----------------------------------------------------------------
	q50          |
	x |   1.003579   .0273909    36.64   0.000     .9498775    1.057281
	_cons |   .0020266   .0138975     0.15   0.884    -.0252202    .0292735
	-------------+----------------------------------------------------------------
	q80          |
	x |   .7244305   .0309551    23.40   0.000     .6637412    .7851198
	_cons |    .398641    .024758    16.10   0.000     .3501016    .4471804
	------------------------------------------------------------------------------
	
	. ldvqreg y_c x, reps(5) q(20 50 80) ll(0) ul(1)
	(running cqr_est on estimation sample)
	
	Bootstrap replications (5)
	----+--- 1 ---+--- 2 ---+--- 3 ---+--- 4 ---+--- 5 
	.....
	
	Censored quantile regression                    Number of obs     =      4,000
	Replications      =          5
	
	------------------------------------------------------------------------------
	|   Observed   Bootstrap                         Normal-based
	y_c |      Coef.   Std. Err.      z    P>|z|     [95
	-------------+----------------------------------------------------------------
	q20          |
	x |   .8210614   .0353826    23.21   0.000     .7517129    .8904099
	_cons |  -.2278362   .0284962    -8.00   0.000    -.2836878   -.1719846
	-------------+----------------------------------------------------------------
	q50          |
	x |   1.000194   .0144165    69.38   0.000      .971938     1.02845
	_cons |     .00578   .0063768     0.91   0.365    -.0067183    .0182782
	-------------+----------------------------------------------------------------
	q80          |
	x |   1.133657   .0676917    16.75   0.000     1.000984    1.266331
	_cons |   .2681894   .0143108    18.74   0.000     .2401407    .2962381
	------------------------------------------------------------------------------
	
\end{stlog}

Note that the coefficients computed by both commands are different. The gap between the two represents the bias for ignoring the censorship process. Figure \ref{fig:figure1} shows the three lines estimated by both commands. The green line correspond to the $\tau=0.9$ case, the blue line to the $\tau=0.5$ one, and finally the red one to $\tau=0.2$. The points with the symbol "\texttt{x}" represent the realizations of the latent (uncensored) variable while the gray circles are those of the observed variable (censored). Clearly, the \texttt{sqreg} command underestimates the coefficients of the extreme quantiles as a consequence of the simulated upper and lower censoring while those estimated by the \texttt{ldvqreg} command are consistent with the scatter of the latent variable.

\begin{figure}
	\caption{Comparing quantile commands under censorship}
	\label{fig:figure2}
	\centering
	\includegraphics[scale=0.33]{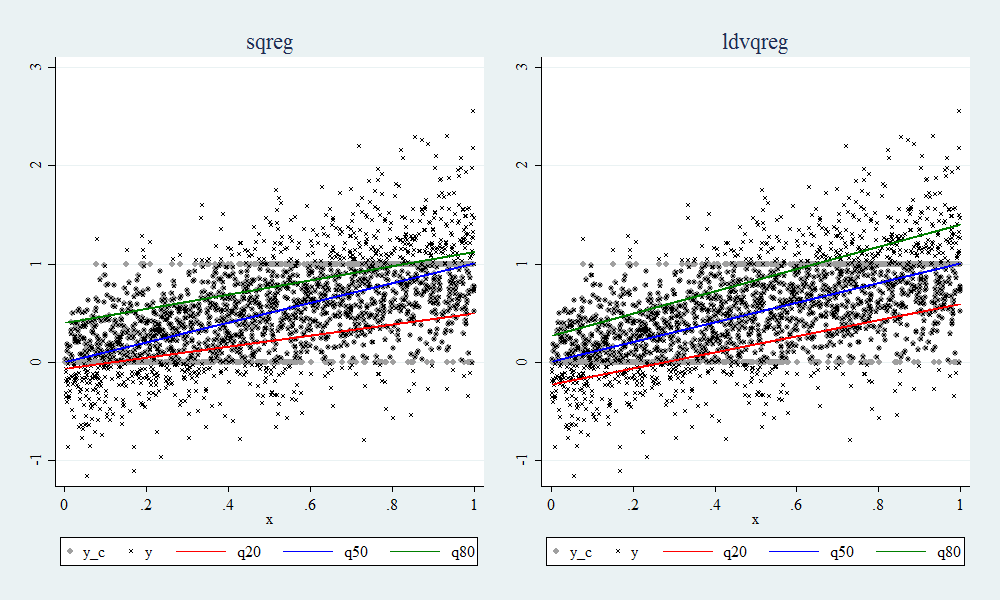}
\end{figure}

\begin{figure}
	\caption{Prediction of quantiles and censorship probability}
	\label{fig:figure3}
	\centering
	\includegraphics[scale=0.33]{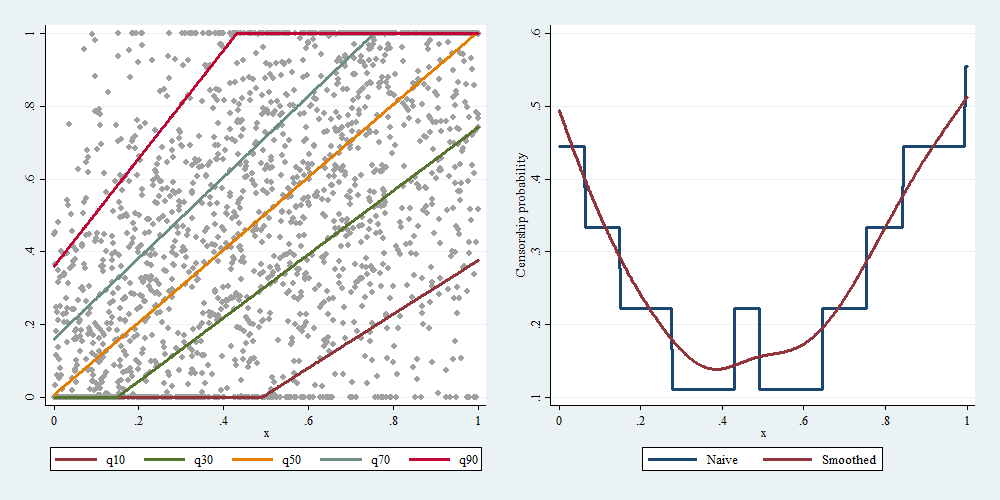}
\end{figure}

The command \texttt{ldvqreg} can be used to predict censored quantiles and also to compute the probability of censorship by the options texttt{qcen()} and \texttt{pcen}, respectively. Quantile prediction can be done in an individual way for each $\tau$, but the probability of censorship requires many $\tau$s (at least 2). We show now an example code and a graph in Figure \ref{fig:figure3} with the predicted censured quantiles (left panel) and probability of censoring (right panel).


\begin{stlog}
	
	. ldvqreg y_c x , reps(2) q(10 20 30 40 50 60 70 80 90) ll(0) ul(1) /*
	*/    qcen(myqcen) pcen(mypcen)
	(running cqr_est on estimation sample)
	
	Bootstrap replications (2)
	----+--- 1 ---+--- 2 ---+--- 3 ---+--- 4 ---+--- 5 
	..
	(output omitted)
	
	. summarize
	
	Variable |        Obs        Mean    Std. Dev.       Min        Max
	-------------+---------------------------------------------------------
	heter |      4,000          .5    .5000625          0          1
	x |      4,000    .4973596    .2883611    .000018   .9997839
	y |      4,000    .5076568    .5295976  -1.163579   2.778667
	y_c |      4,000    .4922552    .3645911          0          1
	mypcen |      4,000      .25675     .125413   .1111111   .5555556
	-------------+---------------------------------------------------------
	mypcen_s |      4,000    .2706427    .1154597   .1384926   .5122733
	myqcen_q10 |      4,000    .0946118    .1214869          0   .3766204
	myqcen_q20 |      4,000    .2127563    .1962473          0   .5930477
	myqcen_q30 |      4,000    .3135726    .2376881          0   .7425613
	myqcen_q40 |      4,000    .4120319    .2696362          0   .8849297
	-------------+---------------------------------------------------------
	myqcen_q50 |      4,000    .5032121    .2883758    .005798          1
	myqcen_q60 |      4,000    .5963741    .3047473   .0502407          1
	myqcen_q70 |      4,000    .6813506    .2779192   .1597887          1
	myqcen_q80 |      4,000    .7620734    .2443681   .2682098          1
	myqcen_q90 |      4,000    .8610955    .2007863   .3594372          1
	
\end{stlog}

It should be noted that in the \texttt{summarize} output there are new variables that were generated with the names used in the \texttt{ldvqreg} run. On one hand, the variables  \texttt{mypcen} and \texttt{mypcen\underline{ }s} have the probability with the na\"ive and the smoothed formulas, respectively. On the other hand, the varlist  \texttt{myqcen\underline{ }q10-myqcen\underline{ }q90} are the predicted censored quantiles  for $\tau\in\{0.10, 0.20, ..., 0.90\}$. 


\subsubsection{Binary dependent variable}

Consider now a binary dependent variable case. The DGP for this case is the following.

\begin{stlog}
	
	. drop _all
	. set seed 321
	. set obs 2000
	number of observations (_N) was 0, now 2,000
	
	. gen x = runiform()*10
	. gen y = -2.5 + x + x*(rchi2(1)-1)/sqrt(2)
	. gen y_b = (y>0)
	
	. ta y_b
	
	y_b |      Freq.     Percent        Cum.
	------------+-----------------------------------
	0 |        883       44.15       44.15
	1 |      1,117       55.85      100.00
	------------+-----------------------------------
	Total |      2,000      100.00
	
\end{stlog}

We first compute a \texttt{probit} model, then a QR model with the (unobserved) latent variable, and finally the proposed BQR using the developed command. For the last two we consider the median estimate, i.e. $\tau=0.5$. In order to compare these two we normalize the coefficients of the median QR regression to $\Vert \boldsymbol{b}\Vert=1$. The binary regression model already has this normalization.

\begin{stlog}
	
	. qui probit y_b x
	
	. nlcom (_b[x]/sqrt(_b[x]^2+_b[_cons]^2)) (_b[_cons]/sqrt(_b[x]^2+_b[_cons]^2))
	
	_nl_1:  _b[x]/sqrt(_b[x]^2+_b[_cons]^2)
	_nl_2:  _b[_cons]/sqrt(_b[x]^2+_b[_cons]^2)
	
	------------------------------------------------------------------------------
	y_b |      Coef.   Std. Err.      z    P>|z|     [95
	-------------+----------------------------------------------------------------
	_nl_1 |   .2230184    .004476    49.83   0.000     .2142456    .2317912
	_nl_2 |  -.9748142    .001024  -951.94   0.000    -.9768213   -.9728072
	------------------------------------------------------------------------------
	
	. 
	. qui bsqreg y  x
	
	. mat norma = e(b)*e(b)'
	
	. gen y_n = y/sqrt(norma[1,1])
	
	. bsqreg y_n x
	(fitting base model)
	
	Bootstrap replications (20)
	----+--- 1 ---+--- 2 ---+--- 3 ---+--- 4 ---+--- 5 
	....................
	
	Median regression, bootstrap(20) SEs                Number of obs =      2,000
	Raw sum of deviations 1378.646 (about .13423963)
	Min sum of deviations 1185.645                    Pseudo R2     =     0.1400
	
	------------------------------------------------------------------------------
	y_n |      Coef.   Std. Err.      t    P>|t|     [95
	-------------+----------------------------------------------------------------
	x |   .2300503   .0081119    28.36   0.000     .2141418    .2459589
	_cons |  -.9731787   .0179613   -54.18   0.000    -1.008404    -.937954
	------------------------------------------------------------------------------
	
	. 
	. ldvqreg y_b x
	(running bqr_est on estimation sample)
	
	Bootstrap replications (20)
	----+--- 1 ---+--- 2 ---+--- 3 ---+--- 4 ---+--- 5 
	....................
	
	Binary quantile regression                      Number of obs     =      2,000
	Replications      =         20
	
	------------------------------------------------------------------------------
	|   Observed   Bootstrap                         Normal-based
	y_b |      Coef.   Std. Err.      z    P>|z|     [95
	-------------+----------------------------------------------------------------
	x |   .2381934   .0069091    34.48   0.000     .2246517     .251735
	_cons |  -.9712218   .0016994  -571.51   0.000    -.9745526   -.9678911
	------------------------------------------------------------------------------
	
\end{stlog}

Note that in this case the BQR model show similar results to the QR model with the latent variable (normalized), but not to the Probit model. In fact, the Probit model should differ from the median estimates given that we are using an asymmetric DGP.


Consider now the comparison of the \texttt{ldvqreg} with the standard QR alternatives in \texttt{STATA}. In this case we compare it with  \texttt{bsqreg} (QR with standard errors computed by bootstrap), separately for different quantiles. 


\begin{stlog}
	
	. bsqreg y_b x, q(20)
	(fitting base model)
	
	Bootstrap replications (20)
	----+--- 1 ---+--- 2 ---+--- 3 ---+--- 4 ---+--- 5 
	.xxxx.x....xx.x..xx.......xx.x.x..
	
	.2 Quantile regression, bootstrap(20) SEs           Number of obs =      2,000
	Raw sum of deviations    223.4 (about 0)
	Min sum of deviations 223.3644                    Pseudo R2     =     0.0002
	
	------------------------------------------------------------------------------
	y_b |      Coef.   Std. Err.      t    P>|t|     [95
	-------------+----------------------------------------------------------------
	x |   .1398933   .0011419   122.51   0.000      .137654    .1421327
	_cons |  -.3986311   .0105947   -37.63   0.000     -.419409   -.3778533
	------------------------------------------------------------------------------
	
	. nlcom (_b[x]/sqrt(_b[x]^2+_b[_cons]^2)) (_b[_cons]/sqrt(_b[x]^2+_b[_cons]^2))
	
	_nl_1:  _b[x]/sqrt(_b[x]^2+_b[_cons]^2)
	_nl_2:  _b[_cons]/sqrt(_b[x]^2+_b[_cons]^2)
	
	------------------------------------------------------------------------------
	y_b |      Coef.   Std. Err.      z    P>|z|     [95
	-------------+----------------------------------------------------------------
	_nl_1 |   .3311357   .0057459    57.63   0.000     .3198739    .3423975
	_nl_2 |  -.9435832   .0020164  -467.94   0.000    -.9475353    -.939631
	------------------------------------------------------------------------------
	
	. ldvqreg y_b x, q(20)
	(running bqr_est on estimation sample)
	
	Bootstrap replications (20)
	----+--- 1 ---+--- 2 ---+--- 3 ---+--- 4 ---+--- 5 
	....................
	
	Binary quantile regression                      Number of obs     =      2,000
	Replications      =         20
	
	------------------------------------------------------------------------------
	|   Observed   Bootstrap                         Normal-based
	y_b |      Coef.   Std. Err.      z    P>|z|     [95
	-------------+----------------------------------------------------------------
	x |   .1410802   .0025301    55.76   0.000     .1361212    .1460391
	_cons |   -.990006   .0003634 -2724.15   0.000    -.9907183   -.9892938
	------------------------------------------------------------------------------
	
	. 
	. bsqreg y_b x, q(50)
	(fitting base model)
	
	Bootstrap replications (20)
	----+--- 1 ---+--- 2 ---+--- 3 ---+--- 4 ---+--- 5 
	....................
	
	Median regression, bootstrap(20) SEs                Number of obs =      2,000
	Raw sum of deviations    441.5 (about 1)
	Min sum of deviations  301.819                    Pseudo R2     =     0.3164
	
	------------------------------------------------------------------------------
	y_b |      Coef.   Std. Err.      t    P>|t|     [95
	-------------+----------------------------------------------------------------
	x |   .1271259   .0011701   108.65   0.000     .1248311    .1294206
	_cons |   -.094313   .0069562   -13.56   0.000    -.1079551   -.0806709
	------------------------------------------------------------------------------
	
	. nlcom (_b[x]/sqrt(_b[x]^2+_b[_cons]^2)) (_b[_cons]/sqrt(_b[x]^2+_b[_cons]^2))
	
	_nl_1:  _b[x]/sqrt(_b[x]^2+_b[_cons]^2)
	_nl_2:  _b[_cons]/sqrt(_b[x]^2+_b[_cons]^2)
	
	------------------------------------------------------------------------------
	y_b |      Coef.   Std. Err.      z    P>|z|     [95
	-------------+----------------------------------------------------------------
	_nl_1 |   .8031167   .0199962    40.16   0.000     .7639249    .8423085
	_nl_2 |  -.5958218   .0269532   -22.11   0.000     -.648649   -.5429946
	------------------------------------------------------------------------------
	
	. ldvqreg y_b x, q(50)
	(running bqr_est on estimation sample)
	
	Bootstrap replications (20)
	----+--- 1 ---+--- 2 ---+--- 3 ---+--- 4 ---+--- 5 
	....................
	
	Binary quantile regression                      Number of obs     =      2,000
	Replications      =         20
	
	------------------------------------------------------------------------------
	|   Observed   Bootstrap                         Normal-based
	y_b |      Coef.   Std. Err.      z    P>|z|     [95
	-------------+----------------------------------------------------------------
	x |   .2381934   .0096565    24.67   0.000     .2192669    .2571198
	_cons |  -.9712218   .0023759  -408.79   0.000    -.9758785   -.9665652
	------------------------------------------------------------------------------
	
	. bsqreg y_b x, q(80)
	(fitting base model)
	convergence not achieved.
	convergence not achieved
	r(430);
	
\end{stlog}

Note that the results are very different across estimators. This applies even if we normalize the \texttt{bsqreg} coefficients to $\Vert \boldsymbol{b}\Vert=1$. In fact, the QR estimates show convergence problems in several bootstrap simulations (denoted by an \texttt{x} in the output), while the \texttt{ldvqreg} runs smoothly. Overall this shows that BQR should be implemented with the proposed smoothed version.


Finally, we implement tests of homogeneity and symmetry, comparing the (true) latent variable model with the binary regression case.  
To evaluate the symmetry of the conditional distribution we use the procedure suggested by \cite{Koenker05} evaluating the following linear null hypothesis: 

\begin{equation*}
H_0: \frac{1}{2}\cdot\boldsymbol{\beta} \left(\frac{1}{2}-\delta \right) + \frac{1}{2}\cdot\boldsymbol{\beta} \left(\frac{1}{2}+\delta \right) - \boldsymbol{\beta} \left(\frac{1}{2}\right) = \boldsymbol{0}     
\end{equation*}
for some $\delta \in (0, \frac{1}{2})$. This can be easily implemented by a Wald test using the \texttt{test} command.

\begin{stlog}
	
	. * With unobservable data (uncensored)
	. sqreg y x , q(10 25 50 75 90) reps(300)
	(fitting base model)
	
	Bootstrap replications (300)
	----+--- 1 ---+--- 2 ---+--- 3 ---+--- 4 ---+--- 5 
	..................................................    50
	..................................................   100
	..................................................   150
	..................................................   200
	..................................................   250
	..................................................   300
	
	Simultaneous quantile regression                    Number of obs =      2,000
	bootstrap(300) SEs                                .10 Pseudo R2 =     0.2544
	.25 Pseudo R2 =     0.1967
	.50 Pseudo R2 =     0.1400
	.75 Pseudo R2 =     0.1630
	.90 Pseudo R2 =     0.2030
	
	------------------------------------------------------------------------------
	|              Bootstrap
	y |      Coef.   Std. Err.      t    P>|t|     [95
	-------------+----------------------------------------------------------------
	q10          |
	x |   .3023696   .0017696   170.86   0.000     .2988991    .3058402
	_cons |  -2.499844   .0022336 -1119.20   0.000    -2.504225   -2.495464
	-------------+----------------------------------------------------------------
	q25          |
	x |   .3563443   .0066943    53.23   0.000     .3432157    .3694729
	_cons |  -2.498919   .0095113  -262.73   0.000    -2.517572   -2.480266
	-------------+----------------------------------------------------------------
	q50          |
	x |   .5799625   .0212302    27.32   0.000     .5383269    .6215981
	_cons |  -2.453407   .0326152   -75.22   0.000    -2.517371   -2.389444
	-------------+----------------------------------------------------------------
	q75          |
	x |   1.174422    .052652    22.31   0.000     1.071164    1.277681
	_cons |  -2.359083   .0911639   -25.88   0.000     -2.53787   -2.180297
	-------------+----------------------------------------------------------------
	q90          |
	x |   2.169029   .0961683    22.55   0.000     1.980428    2.357629
	_cons |   -2.38754   .1410773   -16.92   0.000    -2.664214   -2.110866
	------------------------------------------------------------------------------
	
	. 
	. * Homogeneity
	. test [q10=q25=q50=q75=q90]: x
	
	( 1)  [q10]x - [q25]x = 0
	( 2)  [q10]x - [q50]x = 0
	( 3)  [q10]x - [q75]x = 0
	( 4)  [q10]x - [q90]x = 0
	
	F(  4,  1998) =  118.08
	Prob > F =    0.0000
	
	. 
	. * Symmetry
	. test (([q10]x+[q25]x+[q75]x+[q90]x)/4-[q50]x=0) ///
	>          (([q10]_cons+[q25]_cons+[q75]_cons+[q90]_cons)/4-[q50]_cons=0)
	
	( 1)  .25*[q10]x + .25*[q25]x - [q50]x + .25*[q75]x + .25*[q90]x = 0
	( 2)  .25*[q10]_cons + .25*[q25]_cons - [q50]_cons + .25*[q75]_cons + .25*[q90]_cons = 0
	
	F(  2,  1998) =  176.51
	Prob > F =    0.0000
	
	. 
	. * With observable data (censored)
	. ldvqreg y_b x , q(10 25 50 75 90) reps(300)
	(running bqr_est on estimation sample)
	
	Bootstrap replications (300)
	----+--- 1 ---+--- 2 ---+--- 3 ---+--- 4 ---+--- 5 
	..................................................    50
	..................................................   100
	..................................................   150
	..................................................   200
	..................................................   250
	..................................................   300
	
	Binary quantile regression                      Number of obs     =      2,000
	Replications      =        300
	
	------------------------------------------------------------------------------
	|   Observed   Bootstrap                         Normal-based
	y_b |      Coef.   Std. Err.      z    P>|z|     [95
	-------------+----------------------------------------------------------------
	q10          |
	x |   .1101229   .0532167     2.07   0.039       .00582    .2144257
	_cons |  -.9939236   .0027315  -363.88   0.000    -.9992772   -.9885699
	-------------+----------------------------------------------------------------
	q25          |
	x |   .1548161   .0044306    34.94   0.000     .1461323    .1634998
	_cons |  -.9879507   .0006984 -1414.58   0.000    -.9893196   -.9865819
	-------------+----------------------------------------------------------------
	q50          |
	x |   .2381934   .0096646    24.65   0.000     .2192511    .2571356
	_cons |  -.9712218    .002419  -401.49   0.000    -.9759631   -.9664806
	-------------+----------------------------------------------------------------
	q75          |
	x |   .4386173   .0448135     9.79   0.000     .3507844    .5264501
	_cons |   -.898675   .0230642   -38.96   0.000    -.9438799     -.85347
	-------------+----------------------------------------------------------------
	q90          |
	x |    .722866   .0350865    20.60   0.000     .6540977    .7916342
	_cons |  -.6909895    .035699   -19.36   0.000    -.7609582   -.6210208
	------------------------------------------------------------------------------
	
	. 
	. * Homogeneity
	. test [q10=q25=q50=q75=q90]: x
	
	( 1)  [q10]x - [q25]x = 0
	( 2)  [q10]x - [q50]x = 0
	( 3)  [q10]x - [q75]x = 0
	( 4)  [q10]x - [q90]x = 0
	
	chi2(  4) =  365.80
	Prob > chi2 =    0.0000
	
	. 
	. * Symmetry
	. test (([q10]x+[q25]x+[q75]x+[q90]x)/4-[q50]x=0) ///
	>          (([q10]_cons+[q25]_cons+[q75]_cons+[q90]_cons)/4-[q50]_cons=0)
	
	( 1)  .25*[q10]x + .25*[q25]x - [q50]x + .25*[q75]x + .25*[q90]x = 0
	( 2)  .25*[q10]_cons + .25*[q25]_cons - [q50]_cons + .25*[q75]_cons + .25*[q90]_cons = 0
	
	chi2(  2) =   47.46
	Prob > chi2 =    0.0000
	
\end{stlog}

The results are expected. We reject the hypotheses of both homoscedasticity and symmetry of the latent variable. Both features should indicate that the assumptions of the Probit and Logit models are not valid ad we should consider the semi-parametric approach given by \texttt{ldvqreg}.


Finally, the \texttt{ldvqreg} also computes the conditional probabilities of $y=1$ using the estimated coefficients for a grid of $\tau$s by the option \texttt{p1()}. We show here an example coding:


\begin{stlog}
	
	. ldvqreg y_b x , reps(2) q(10 20 30 40 50 60 70 80 90) ll(0) ul(1) p1(p_bqr)
	(running bqr_est on estimation sample)
	
	Bootstrap replications (2)
	----+--- 1 ---+--- 2 ---+--- 3 ---+--- 4 ---+--- 5 
	..
	(output omitted)
	
	. probit y_b x
	
	Iteration 0:   log likelihood =  -1372.574  
	Iteration 1:   log likelihood =  -922.7862  
	Iteration 2:   log likelihood = -921.47418  
	Iteration 3:   log likelihood = -921.47373  
	Iteration 4:   log likelihood = -921.47373  
	
	Probit regression                               Number of obs     =      2,000
	LR chi2(1)        =     902.20
	Prob > chi2       =     0.0000
	Log likelihood = -921.47373                     Pseudo R2         =     0.3287
	
	------------------------------------------------------------------------------
	y_b |      Coef.   Std. Err.      z    P>|z|     [95
	-------------+----------------------------------------------------------------
	x |   .3639832   .0141984    25.64   0.000     .3361549    .3918116
	_cons |  -1.590972   .0738312   -21.55   0.000    -1.735679   -1.446265
	------------------------------------------------------------------------------
	
	. predict p_pro
	(option pr assumed; Pr(y_b))
	
	. summarize
	
	Variable |        Obs        Mean    Std. Dev.       Min        Max
	-------------+---------------------------------------------------------
	x |      2,000      4.9855    2.852372   .0001804    9.99784
	y |      2,000    2.408413    6.126037  -2.499892   60.79322
	y_b |      2,000       .5585    .4966901          0          1
	y_n |      2,000    .9553311     2.42998  -.9916178   24.11449
	p_bqr |      2,000    .5653889    .3212985          0          1
	-------------+---------------------------------------------------------
	p_bqr_s |      2,000    .5661044    .3177954   2.72e-13   .9842031
	p_pro |      2,000    .5578556    .3137343   .0558153   .9797236
	
\end{stlog}

\begin{figure}
	\caption{Comparison of predicted probabilities}
	\label{fig:figure4}
	\centering
	\includegraphics[scale=0.33]{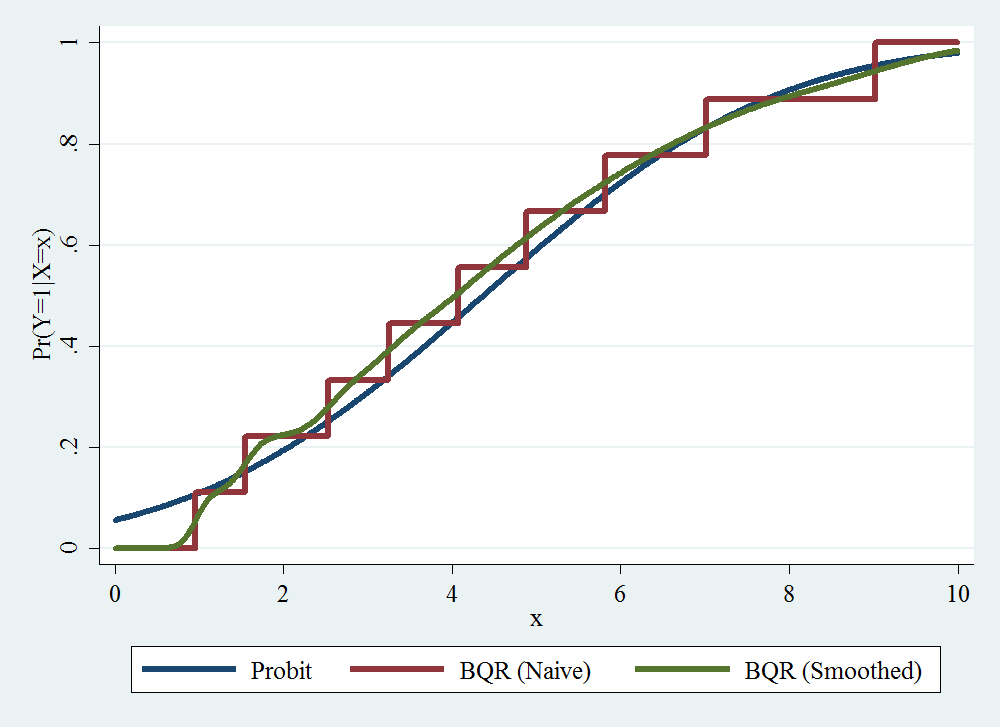}
\end{figure}

Note that new variables appear, that is,  \texttt{p\underline{ }bqr} and \texttt{p\underline{ }bqr} generated by  \texttt{ldvqreg} and the variable \texttt{p\underline{ }pro} generated by  \texttt{probit}. Figure \ref{fig:figure4} shows these three predicted probabilities of $y=1|\boldsymbol{x}$. It should be noted that the Probit model underestimates the probabilities for the center values and overestimates in the extremes, in particular for low \texttt{x}. This is an expected result because of the heterogeneity and asymmetries in the DGP.


\subsection{Example 2: Labor Supply Models}

In this section we show an example of the \texttt{ldvqreg} command applied to the study of hours worked and the probability of having a job in Uruguay. The data comes from the 2015 Continuous Household Survey (ECH for its acronym in Spanish) prepared by the National Institute of Statistics (INE) and the sample consists of women between 18 and 45 years old who are in the labor force and living in urban areas of Montevideo. It is usual in the literature to treat hours worked as a censored variable and therefore to use a Tobit type I model in empirical studies. On the other hand, binary Probit/Logit models are widely used to study the probability of having a job. In both cases, the \texttt{ldvqreg} command is a flexible alternative that allows evaluating some key assumptions of the mentioned maximum likelihood models such as homoscedasticity and symmetry of the conditional distribution.

First we show the model for hours worked. The covariates are age, years of education, the number of children under 6 years of age in the household, and three dummy variables that indicate whether the woman is married, whether she is the head of the household, and whether her partner has a job. Below is the code with a description of the variables together with the result of the \texttt{tobit} and \texttt{ldvqreg} commands applied to the 0.20, 0.50 and 0.80 quantiles. For clarity of exposition, the coefficients are also shown in Table \ref{table1}.

\begin{stlog}
	. use labordatauy, clear
	. describe
	
	Contains data from labordatauy.dta
	obs:         9,601                          
	vars:             8                          19 Sep 2021 10:01
	size:       307,232                          
	--------------------------------------------------------------------------------------
	storage   display    value
	variable name   type    format     label      variable label
	--------------------------------------------------------------------------------------
	work            float   
	hours           float   
	age             float   
	educ            float   
	married         float   
	children        float   
	couplewrk       float   
	head            float   
	--------------------------------------------------------------------------------------
	
	. tobit hours age educ married children couplewrk head, ll(0)
	
	Tobit regression                                Number of obs     =      9,601
	LR chi2(6)        =     453.89
	Prob > chi2       =     0.0000
	Log likelihood = -38264.287                     Pseudo R2         =     0.0059
	
	------------------------------------------------------------------------------
	hours |      Coef.   Std. Err.      t    P>|t|     [95
	-------------+----------------------------------------------------------------
	age |   .3345651   .0269592    12.41   0.000     .2817195    .3874108
	educ |   .4935618   .0514794     9.59   0.000     .3926513    .5944724
	married |   2.646408   .5473636     4.83   0.000      1.57346    3.719357
	children |  -2.297002   .3264407    -7.04   0.000    -2.936895    -1.65711
	couplewrk |  -1.529182    .534851    -2.86   0.004    -2.577603   -.4807612
	head |   2.032225   .4190952     4.85   0.000      1.21071    2.853741
	_cons |   13.67016   1.034941    13.21   0.000     11.64146    15.69887
	-------------+----------------------------------------------------------------
	/sigma |   18.11237   .1431627                      17.83174      18.393
	------------------------------------------------------------------------------
	1,037  left-censored observations at hours <= 0
	8,564     uncensored observations
	0 right-censored observations
	
	. outreg2 using "cen-qreg.xls", replace
	cen-qreg.xls
	dir : seeout
	
	. 
	. ldvqreg hours age educ married children couplewrk head, ll(0) reps(100) q(20 50 80)
	(running cqr_est on estimation sample)
	
	Bootstrap replications (100)
	----+--- 1 ---+--- 2 ---+--- 3 ---+--- 4 ---+--- 5 
	..................................................    50
	..................................................   100
	
	Censored quantile regression                    Number of obs     =      9,601
	Replications      =        100
	
	------------------------------------------------------------------------------
	|   Observed   Bootstrap                         Normal-based
	hours |      Coef.   Std. Err.      z    P>|z|     [95
	-------------+----------------------------------------------------------------
	q20          |
	age |   .5395249   .0504014    10.70   0.000     .4407399    .6383099
	educ |   1.617458    .078667    20.56   0.000     1.463273    1.771642
	married |   4.504101   .9645752     4.67   0.000     2.613568    6.394634
	children |  -3.008667   .5073226    -5.93   0.000    -4.003001   -2.014333
	couplewrk |  -2.672318   .9351631    -2.86   0.004    -4.505204   -.8394319
	head |   3.875635   .7831536     4.95   0.000     2.340682    5.410588
	_cons |  -22.81325   1.393761   -16.37   0.000    -25.54497   -20.08153
	-------------+----------------------------------------------------------------
	q50          |
	age |   .1841254   .0296005     6.22   0.000     .1261096    .2421413
	educ |   .0484223   .0603246     0.80   0.422    -.0698118    .1666564
	married |   3.138284   .7572092     4.14   0.000     1.654181    4.622386
	children |  -2.693533   .4289794    -6.28   0.000    -3.534318   -1.852749
	couplewrk |  -2.414126   .6813139    -3.54   0.000    -3.749476   -1.078775
	head |   1.058981   .3585019     2.95   0.003     .3563303    1.761632
	_cons |   29.86262   1.365012    21.88   0.000     27.18724    32.53799
	-------------+----------------------------------------------------------------
	q80          |
	age |   .0454839   .0129672     3.51   0.000     .0200687    .0708991
	educ |   -.488003   .0267038   -18.27   0.000    -.5403416   -.4356645
	married |   .4958638   .2746884     1.81   0.071    -.0425155    1.034243
	children |  -.5143125   .1713502    -3.00   0.003    -.8501528   -.1784723
	couplewrk |  -.5283063   .2773909    -1.90   0.057    -1.071983    .0153699
	head |   .3686784   .2151083     1.71   0.087    -.0529261    .7902829
	_cons |   48.86151   .4881569   100.09   0.000     47.90474    49.81828
	------------------------------------------------------------------------------
	
	. outreg2 using "cen-qreg.xls", append
	cen-qreg.xls
	dir : seeout
	
	. test [q20=q50=q80]
	
	( 1)  [q20]age - [q50]age = 0
	( 2)  [q20]educ - [q50]educ = 0
	( 3)  [q20]married - [q50]married = 0
	( 4)  [q20]children - [q50]children = 0
	( 5)  [q20]couplewrk - [q50]couplewrk = 0
	( 6)  [q20]head - [q50]head = 0
	( 7)  [q20]age - [q80]age = 0
	( 8)  [q20]educ - [q80]educ = 0
	( 9)  [q20]married - [q80]married = 0
	(10)  [q20]children - [q80]children = 0
	(11)  [q20]couplewrk - [q80]couplewrk = 0
	(12)  [q20]head - [q80]head = 0
	
	chi2( 12) = 2924.46
	Prob > chi2 =    0.0000
	
	. test (0.5*_b[q20:age]+0.5*_b[q80:age]=_b[q50:age]) ///
	>          (0.5*_b[q20:edu]+0.5*_b[q80:edu]=_b[q50:edu]) ///      
	>          (0.5*_b[q20:mar]+0.5*_b[q80:mar]=_b[q50:mar]) ///      
	>          (0.5*_b[q20:chi]+0.5*_b[q80:chi]=_b[q50:chi]) ///      
	>          (0.5*_b[q20:cou]+0.5*_b[q80:cou]=_b[q50:cou]) ///      
	>          (0.5*_b[q20:hea]+0.5*_b[q80:hea]=_b[q50:hea]) ///      
	>          (0.5*_b[q20:_co]+0.5*_b[q80:_co]=_b[q50:_co])  
	
	( 1)  .5*[q20]age - [q50]age + .5*[q80]age = 0
	( 2)  .5*[q20]educ - [q50]educ + .5*[q80]educ = 0
	( 3)  .5*[q20]married - [q50]married + .5*[q80]married = 0
	( 4)  .5*[q20]children - [q50]children + .5*[q80]children = 0
	( 5)  .5*[q20]couplewrk - [q50]couplewrk + .5*[q80]couplewrk = 0
	( 6)  .5*[q20]head - [q50]head + .5*[q80]head = 0
	( 7)  .5*[q20]_cons - [q50]_cons + .5*[q80]_cons = 0
	
	chi2(  7) = 1054.47
	Prob > chi2 =    0.0000
	
\end{stlog}

\begin{table}
	\caption{Working hours model}\label{table1}
	\centering
	
	\begin{tabular}{rrccc}
		\hline
		{\bf } &     {\bf } & \multicolumn{ 3}{c}{{\bf Censored QR}} \\
		
		{\bf } & {\bf Tobit} &  {\bf q20} &  {\bf q50} &  {\bf q80} \\
		\hline
		Age &   0.335*** &   0.540*** &   0.184*** &  0.0455*** \\
		
		&   (0.0270) &   (0.0504) &   (0.0296) &   (0.0130) \\
		
		Years of education &   0.494*** &   1.617*** &     0.0484 &  -0.488*** \\
		
		&   (0.0515) &   (0.0787) &   (0.0603) &   (0.0267) \\
		
		Married &   2.646*** &   4.504*** &   3.138*** &     0.496* \\
		
		&    (0.547) &    (0.965) &    (0.757) &    (0.275) \\
		
		Children under 6 yrs &  -2.297*** &  -3.009*** &  -2.694*** &  -0.514*** \\
		
		&    (0.326) &    (0.507) &    (0.429) &    (0.171) \\
		
		Couple working &  -1.529*** &  -2.672*** &  -2.414*** &    -0.528* \\
		
		&    (0.535) &    (0.935) &    (0.681) &    (0.277) \\
		
		Household Head &   2.032*** &   3.876*** &   1.059*** &     0.369* \\
		
		&    (0.419) &    (0.783) &    (0.359) &    (0.215) \\
		
		Constant &   13.67*** &  -22.81*** &   29.86*** &   48.86*** \\
		
		&    (1.035) &    (1.394) &    (1.365) &    (0.488) \\
		\hline
		Observations &      9,601 &      9,601 &      9,601 &      9,601 \\
		\hline
	\end{tabular}  
	
	\footnotesize
	Source: own estimates based on the ECH 2015 (INE). 
	Note: standard errors in parentheses, * indicates significance at 10 \%, ** at 5 \% and *** at 1 \%.
\end{table}  

Note that the tests show that homoscedasticity and symmetry of the conditional distribution are rejected at the usual levels of significance. Therefore, the overall effect of the covariates is heterogeneous and it is also likely that the coefficients estimated by Tobit model are potentially biased.

Second, we analyze the binary regression model using the variable \texttt{work} as a dependent variable, which is a dummy variable that indicates with 1 if the woman is working and 0 otherwise. As in the previous example, we start by showing the result of the Probit model where we normalize the coefficients in such a way that $\Vert \boldsymbol{b}\Vert=1$ and therefore are comparable with those of the output of the \texttt{ldvqreg} command. Again, we estimate the parameters of the conditional quantiles 0.20, 0.50 and 0.80 of the latent variable. The code is shown below and the results, for simplicity, are shown in Table \ref{table2}:

\begin{stlog}
	
	. probit work age educ married children couplewrk head
	
	Iteration 0:   log likelihood = -3286.7444  
	Iteration 1:   log likelihood = -2928.7265  
	Iteration 2:   log likelihood = -2917.8391  
	Iteration 3:   log likelihood = -2917.8201  
	Iteration 4:   log likelihood = -2917.8201  
	
	Probit regression                               Number of obs     =      9,601
	LR chi2(6)        =     737.85
	Prob > chi2       =     0.0000
	Log likelihood = -2917.8201                     Pseudo R2         =     0.1122
	
	------------------------------------------------------------------------------
	work |      Coef.   Std. Err.      z    P>|z|     [95
	-------------+----------------------------------------------------------------
	age |   .0402997   .0026915    14.97   0.000     .0350244     .045575
	educ |   .0831836    .005493    15.14   0.000     .0724175    .0939497
	married |   .2323897   .0494435     4.70   0.000     .1354823    .3292971
	children |  -.0802435   .0321561    -2.50   0.013    -.1432682   -.0172187
	couplewrk |   -.051575   .0468705    -1.10   0.271    -.1434395    .0402895
	head |   .2070765   .0440624     4.70   0.000     .1207159    .2934372
	_cons |  -1.035893   .1003225   -10.33   0.000    -1.232521   -.8392642
	------------------------------------------------------------------------------
	
	. mat b_probit = e(b)
	
	. svmat b_probit
	
	. sca norm = 0
	
	. foreach b of varlist b_probit* {
		2.                 sca norm = norm + `b'[1]^2
		3.         }
	
	. sca norm = sqrt(norm)
	
	. outreg2 using "bin-qreg.xls", stnum(replace coef=coef/norm, replace se=se/norm) replace
	bin-qreg.xls
	dir : seeout
	
	. ldvqreg work age educ married children couplewrk head, reps(100) q(20 50 80)
	(running bqr_est on estimation sample)
	
	Bootstrap replications (100)
	----+--- 1 ---+--- 2 ---+--- 3 ---+--- 4 ---+--- 5 
	..................................................    50
	..................................................   100
	
	Binary quantile regression                      Number of obs     =      9,601
	Replications      =        100
	
	------------------------------------------------------------------------------
	|   Observed   Bootstrap                         Normal-based
	work |      Coef.   Std. Err.      z    P>|z|     [95
	-------------+----------------------------------------------------------------
	q20          |
	age |   .0292285   .0035849     8.15   0.000     .0222022    .0362549
	educ |   .0323487   .0085367     3.79   0.000      .015617    .0490803
	married |   .2048792   .0422928     4.84   0.000     .1219869    .2877715
	children |  -.0749141   .0393592    -1.90   0.057    -.1520567    .0022285
	couplewrk |  -.1221973   .0423323    -2.89   0.004    -.2051671   -.0392276
	head |   .1493376   .0565451     2.64   0.008     .0385111     .260164
	_cons |  -.9556585   .0162207   -58.92   0.000    -.9874504   -.9238665
	-------------+----------------------------------------------------------------
	q50          |
	age |   .0257198   .0232722     1.11   0.269    -.0198928    .0713324
	educ |   .1716362   .1126731     1.52   0.128    -.0491991    .3924715
	married |   .4773294   .2691138     1.77   0.076    -.0501239    1.004783
	children |    .117616   .1788098     0.66   0.511    -.2328448    .4680768
	couplewrk |  -.1989163   .2876677    -0.69   0.489    -.7627346     .364902
	head |   .2621181   .1656298     1.58   0.114    -.0625104    .5867466
	_cons |  -.7873564   .1550116    -5.08   0.000    -1.091174   -.4835392
	-------------+----------------------------------------------------------------
	q80          |
	age |   .0720622   .0227476     3.17   0.002     .0274777    .1166467
	educ |   .0672349   .0912818     0.74   0.461    -.1116741    .2461438
	married |   .2004946   .1340388     1.50   0.135    -.0622167    .4632059
	children |  -.0696732   .1128843    -0.62   0.537    -.2909224     .151576
	couplewrk |  -.0327274   .2034881    -0.16   0.872    -.4315568     .366102
	head |   .1811072   .0872446     2.08   0.038      .010111    .3521035
	_cons |  -.9546537   .1201494    -7.95   0.000    -1.190142   -.7191653
	------------------------------------------------------------------------------
	
	. outreg2 using "bin-qreg.xls", append
	bin-qreg.xls
	dir : seeout
	
	. test [q20=q50=q80]
	
	( 1)  [q20]age - [q50]age = 0
	( 2)  [q20]educ - [q50]educ = 0
	( 3)  [q20]married - [q50]married = 0
	( 4)  [q20]children - [q50]children = 0
	( 5)  [q20]couplewrk - [q50]couplewrk = 0
	( 6)  [q20]head - [q50]head = 0
	( 7)  [q20]age - [q80]age = 0
	( 8)  [q20]educ - [q80]educ = 0
	( 9)  [q20]married - [q80]married = 0
	(10)  [q20]children - [q80]children = 0
	(11)  [q20]couplewrk - [q80]couplewrk = 0
	(12)  [q20]head - [q80]head = 0
	
	chi2( 12) =   24.58
	Prob > chi2 =    0.0169
	
	. test (0.5*_b[q20:age]+0.5*_b[q80:age]=_b[q50:age]) ///
	>          (0.5*_b[q20:edu]+0.5*_b[q80:edu]=_b[q50:edu]) ///      
	>          (0.5*_b[q20:mar]+0.5*_b[q80:mar]=_b[q50:mar]) ///      
	>          (0.5*_b[q20:chi]+0.5*_b[q80:chi]=_b[q50:chi]) ///      
	>          (0.5*_b[q20:cou]+0.5*_b[q80:cou]=_b[q50:cou]) ///      
	>          (0.5*_b[q20:hea]+0.5*_b[q80:hea]=_b[q50:hea]) ///      
	>          (0.5*_b[q20:_co]+0.5*_b[q80:_co]=_b[q50:_co])  
	
	( 1)  .5*[q20]age - [q50]age + .5*[q80]age = 0
	( 2)  .5*[q20]educ - [q50]educ + .5*[q80]educ = 0
	( 3)  .5*[q20]married - [q50]married + .5*[q80]married = 0
	( 4)  .5*[q20]children - [q50]children + .5*[q80]children = 0
	( 5)  .5*[q20]couplewrk - [q50]couplewrk + .5*[q80]couplewrk = 0
	( 6)  .5*[q20]head - [q50]head + .5*[q80]head = 0
	( 7)  .5*[q20]_cons - [q50]_cons + .5*[q80]_cons = 0
	
	chi2(  7) =    3.45
	Prob > chi2 =    0.8406
	
\end{stlog}

\begin{table}
	\caption{Probability of having a job}\label{table2}
	\centering
	
	\begin{tabular}{rrccc}
		\hline
		{\bf } &     {\bf } & \multicolumn{ 3}{c}{{\bf Binary QR}} \\
		
		{\bf } & {\bf Probit} &  {\bf q20} &  {\bf q50} &  {\bf q80} \\
		\hline
		Age &  0.0370*** &  0.0292*** &     0.0257 &  0.0721*** \\
		
		&  (0.00247) &  (0.00358) &   (0.0233) &   (0.0227) \\
		
		Years of education &  0.0763*** &  0.0323*** &      0.172 &     0.0672 \\
		
		&  (0.00504) &  (0.00854) &    (0.113) &   (0.0913) \\
		
		Married &   0.213*** &   0.205*** &     0.477* &      0.200 \\
		
		&   (0.0454) &   (0.0423) &    (0.269) &    (0.134) \\
		
		Children under 6 yrs &  -0.0736** &   -0.0749* &      0.118 &    -0.0697 \\
		
		&   (0.0295) &   (0.0394) &    (0.179) &    (0.113) \\
		
		Couple working &    -0.0473 &  -0.122*** &     -0.199 &    -0.0327 \\
		
		&   (0.0430) &   (0.0423) &    (0.288) &    (0.203) \\
		
		Household Head &   0.190*** &   0.149*** &      0.262 &    0.181** \\
		
		&   (0.0404) &   (0.0565) &    (0.166) &   (0.0872) \\
		
		Constant &  -0.951*** &  -0.956*** &  -0.787*** &  -0.955*** \\
		
		&   (0.0921) &   (0.0162) &    (0.155) &    (0.120) \\
		\hline
		Observations &      9,601 &      9,601 &      9,601 &      9,601 \\
		\hline
	\end{tabular}  
	
	\footnotesize
	Source: own estimates based on the ECH 2015 (INE).\\
	Note: standard errors in parentheses, * indicates significance at 10 \%, ** at 5 \% and *** at 1 \%; all coefficients are normalized such that $\Vert \boldsymbol{b} \Vert=1$.
	
\end{table}  

In this case it is not possible to reject neither the hypotheses of symmetry nor homoscedasticity at 1\% significance, so it seems that a standard model like the Probit works relatively well. Although the latter gives the impression that the \texttt{ldvqreg} command is somewhat innocuous, it should be noted that it serves either as an empirical way of validating or refuting (and possibly replacing) the estimates of a Probit/Logit model.

\section{Comments and suggestions} \label{conclusion}

This paper proposes a new \texttt{Stata} command \texttt{ldvqreg} to estimate censored quantile regression and binary regression models. A key feature of the proposed command is that it implements a smoothed version of the quantile regression model. Thus, it works very well for the case of censoring and binary dependent variable.

We illustrate the potential pitfalls of ignoring the censoring mechanisms. In the simulation exercises we compare the standard quantile regression estimates with our corrected procedure. The results highlight that the former may be biased in the usual way. In fact this is the same issue that may appear if we compare the Tobit model with standard OLS. The same applies to the binary regression model. In this case, we show that the Probit model may be biased in the underlying data generating process is not homoscedastic and symmetric, while the implementation using standard quantile regression estimates may suffer from convergence problems. In all cases the command  \texttt{ldvqreg} clearly solves this issues.

\bibliographystyle{ecta}
\bibliography{ldvqreg}

\end{document}